\documentclass[a4paper,11pt]{article}            
\usepackage{cite}
\usepackage{graphicx}
\usepackage{float}
\usepackage{wrapfig}
\usepackage{authblk}
\newcommand{\del}{\partial}
\begin{document}

\title{Speed of transverse waves in a string revisited}
\author{Joseph A Rizcallah\\joeriz68@gmail.com}
\affil{School of Education, Lebanese University, Beirut, Lebanon}
\date{}
\maketitle

\begin{abstract}
In many introductory-level physics textbooks, the derivation of the formula 
for the speed of transverse waves in a string is either omitted altogether or presented under physically overly idealized assumptions about the shape of the considered wave pulse and the related velocity and acceleration distributions. In the paper, we derive the named formula by applying Newton’s second law or the work-energy theorem to a finite element of the string, making no assumptions about the shape of the wave. We argue that the suggested method can help the student gain a deeper insight into the nature of waves and the related process of energy transport, as well as provide a new experience with the fundamental principles of mechanics as applied to extended and deformable bodies.
\end{abstract}

\section{Introduction}\label{intro}
Several algebra-based treatments~\cite{gian,giam} state and discuss the formula 
\begin{equation}\label{speed}
c=\sqrt{\frac{T}{\mu}},
\end{equation}
for the speed $c$ of a transverse wave in a string, with constant linear mass density $\mu$ and under uniform tension $T$, but fail to derive it. On the other hand, to obtain formula (\ref{speed}), standard calculus-based physics textbooks~\cite{young, tip, gianc} apply Newton's second law to a ramp-shaped pulse or employ the wave equation, the latter with the stipulation that the student has some background in multivariable calculus. An exception is the textbook by Serway and Jewet~\cite{ser}, where (\ref{speed}) is arrived at by transforming the problem into the co-moving frame of the wave and then applying Newton's second law to an infinitesimal string element. Another, almost entirely algebraic, treatment that also applies Newton's second law to an infinitesimal element of the string is offered in~\cite{sob}.

Although the ramp-shaped pulse makes (\ref{speed}) a fairly straightforward consequence of Newton's second law (in impulse-momentum form) for a variable mass system, it leaves the question open as to whether (\ref{speed}) holds for a wave of arbitrary shape. Moreover, the ideal ramp-shaped pulse has an inherent infinity associated with it. Namely, the abrupt change in the transverse velocity of the string at the leading edge of the ramp implies infinite acceleration of the corresponding element of the string. To avoid the infinite acceleration, of course, some curvature of the string should be introduced at the leading edge. This, however, would spoil the ramp shape and render the impulse-momentum derivation valid only for sufficiently late times (after the string has been first disturbed).

Furthermore, the method of~\cite{ser} avoids dealing with the speed of propagation upfront. Instead, it does away with the motion of the wave and applies Newton's second law to an infinitesimal element of the string in its apparent “sliding” motion. While this clever trick quickly leads to the desired result, pedagogically it overlooks the opportunity of reinforcing the principles of Newtonian mechanics as applied to the extended and deformable string. Likewise, in the neat approach of~\cite{sob}, Newton's second law is also applied to an infinitesimal, rather than finite, element of the string oscillating in its fundamental mode (essentially a harmonic oscillator).

In what follows, after we briefly introduce our notation and establish some well-known results, we employ Newton's second law and the work-energy theorem for a finite element of the string to present two alternative derivations of (\ref{speed}). The suggested approach bears some resemblance to the derivation of the continuity and Bernoulli equations in fluid dynamics as commonly presented in introductory physics textbooks, which makes it accessible as well as instructive for the students.

\section{Notation and auxiliary results}
\begin{wrapfigure}{r}{0.5\textwidth}
\begin{center}
\includegraphics[scale=0.5]{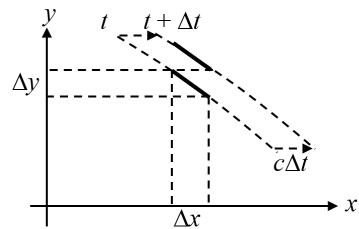}
\end{center}
\caption{A line segment displaced “upward” by a wave propagating to the right at speed $c$.}
\label{fig1}
\end{wrapfigure}

The wave function $y(x,t)=f(x-ct)$, where $c$ is the wave speed, describes a non-dispersive wave propagating down a string stretched along the $x$-axis, which we shall take as the string's equilibrium state. This functional dependence of $f$ on $x$ and $t$ implies that the wave maintains its shape as it moves “rigidly” to the right (the positive $x$-direction). A simple, yet useful, consequence of this fact is the well-known relation between the local transverse velocity $v$ of the string and its slope $s$
\begin{equation}\label{speedslope}
v=-cs.
\end{equation}
To establish (\ref{speedslope}), we consider a string element of length so small (infinitesimal in the limit) that it can be treated as a straight line segment, as shown in figure~\ref{fig1}. The element has height $|\Delta y|$ and negative slope $s=-\frac{|\Delta y|}{\Delta x}$. As the wave advances a distance $\Delta x$ to the right, the element's right edge undergoes an “upward” displacement equal in magnitude to its height $|\Delta y|$. Therefore, the transverse velocity of the element (precisely, of its right edge) is $v=\frac{|\Delta y|}{\Delta t}$, which together with $c=\frac{\Delta x}{\Delta t}$ and $s=-\frac{|\Delta y|}{\Delta x}$ yields (\ref{speedslope}). For a string element with a positive angle of tilt, one instead has $v=-\frac{|\Delta y|}{\Delta t}$ and $s=\frac{|\Delta y|}{\Delta x}$, yet (\ref{speedslope}) continues to hold. (Strictly speaking, equation (\ref{speedslope}) is exact only in the limit $x \rightarrow 0$ ($t \rightarrow 0$), i.e. for the partial derivatives $\frac{\del f}{\del t}$ and $\frac{\del f}{\del x}$, being a consequence of the dependence of the function $f$ on $x$ and $t$ via the wave argument~$x-ct$.)

We now turn to the relation between the transverse force $F$ that acts at a given point of the string and the slope $s$ at that point. As is the custom, we make the following simplifying assumptions: (i) the string undergoes no longitudinal displacement, (ii) that it is perfectly flexible, so that the tension $ \vec{T}$ at any point acts along the local tangent, and (iii) that the wave is so small in amplitude that the approximate equality $\sin \theta\approx\tan \theta\approx\theta$ holds everywhere for the local angle of tilt $\theta$. A well-known consequence of these assumptions is the uniformity of the tension $T$ along the string. 

\begin{wrapfigure}{r}{0.5\textwidth}
\begin{center}
\includegraphics[scale=0.5]{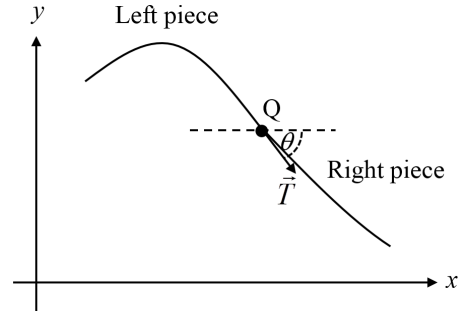}
\end{center}
\caption{The force $\vec{T}$ exerted at point Q by the right piece of the string on the left one. The transverse component of this force is $F=Ts$.}
\label{fig2}
\end{wrapfigure}
To find $F$, consider an imaginary cut made in the string at an arbitrary point Q, as shown in figure~\ref{fig2}. The cut splits the string into two pieces: left and right. The force $\vec{T}$ exerted by the right piece on the left piece is depicted in figure~\ref{fig2}. Its transverse component $F$ is given by $F=T\sin\theta$, which together with the approximation $\sin \theta \approx \tan \theta \equiv s$ yields $F=Ts$. By Newton's third law, the force exerted by the left piece on the right one has a transverse component $F=-Ts$. Hence, the transverse force that acts at a point is given by 
\begin{equation}\label{force}
F=\pm Ts,
\end{equation}
where the plus or minus sign is used depending on whether the force acts on the left or on the right piece. As an aside, note that the form of the wave function $y(x,t)=f(x-ct)$ has not been used in obtaining (\ref{force}), so the latter holds for arbitrary small deformations of a flexible string, including static ones.  

Finally, we need to obtain an expression for the potential energy associated with a small string element. Needless to say, the potential energy we have in mind here is due to the deformation of the string itself rather than its interaction with, say, gravitational or electric fields. To establish such an expression, we again consider a small string element as the one depicted in figure~\ref{fig1}. In the equilibrium state the element is horizontal and has length $\Delta x$. In the perturbed state (figure~\ref{fig1}), however, the element is tilted through angle $\theta$ and its length, by the Pythagorean theorem, increases to $\Delta x \sqrt{1+\tan^2 \theta} = \Delta x \sqrt{1+s^2} \approx \Delta x (1+s^2/2)$, where in the last approximate equality the binomial expansion has been used. This elongation takes place under constant tension $T$ and so (by analogy with stretching a spring) involves an amount of work equal to $T(\Delta x (1+s^2/2) - \Delta x) = 1/2Ts^2\Delta x$ being done on the string. Since this work is stored in the string and can be entirely retrieved (ignoring hysteresis effects), it can be associated with some kind of elastic potential energy $U$ of deformation, defined up to an arbitrary additive constant dependent on the choice of reference~\cite{king,math,burko}:
\begin{equation}\label{energy}
U=1/2Ts^2\Delta x.
\end{equation}
In passing, we note that (\ref{energy}) has nothing to do with the motion (in particular wave motion) of the string and that it holds true for arbitrary small deformations of a flexible string. At this point the scene is set and we now proceed to the derivation.

\section{Derivation of equation (\ref{speed})} 
\begin{figure}[H]
\centering
\includegraphics[scale=0.5]{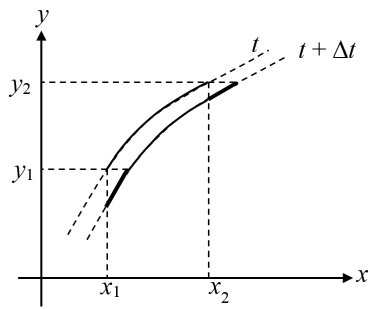}
\caption{A stretch of string between $x_1$ and $x_2$. The displacement of its CM is due to the displacements at its ends marked by heavy segments.}
\label{fig3}
\end{figure}

Consider a finite stretch of string whose left and right ends are at $x_1$ and $x_2 > x_1$ respectively, as shown by the solid line in figure~\ref{fig3}. The mass of the string between $x_1$ and $x_2$ is $\mu(x_2-x_1)$, where $\mu$ is the string's linear mass density, assumed uniform. Let $y_1$ and $y_2$ denote the transverse displacements of the string at the ends at some time $t$. 
\label{der}
From figure~\ref{fig3} we see that, with the advance of the wave, the displacements as well as the slopes are merely transmitted to the right from one element to another everywhere within the considered stretch except at its ends, where new values of these quantities “enter” the stretch through $x_1$ and old ones “leave” it through $x_2$. As we shall see below, this observation is crucial for our purposes.

We want to apply Newton's second law to the stretch of string between $x_1$ and $x_2$, more precisely to its center of mass (CM). To this end, picture the string between $x_1$ and $x_2$ partitioned into very small (infinitesimal in the limit) elements, each of length $\Delta x$ and mass $\mu \Delta x$. In a time $\Delta t = \Delta x/c$, during which the wave advances $\Delta x$ to the right, the center of mass undergoes a transverse displacement $\Delta y_{CM}$ equal to
\begin{equation}\label{cm}
\Delta y_{CM}=\frac{(\mu\Delta x)y_1-(\mu\Delta x)y_2}{\mu(x_2-x_1)}=\frac{y_1-y_2}{x_2-x_1}\Delta x.
\end{equation}
To see why equation (\ref{cm}) holds, suffice it to recall the definition of $y_{CM}$ as the sum of terms of the form $(\mu \Delta x)y$ for each of the elements constituting the stretch of string and note that, with the advance of the wave, only the end elements at $x_1$ and $x_2$ contribute to the change in $y_{CM}$ (see figure~\ref{fig3}), whereas the $(\mu \Delta x)y$ terms corresponding to the interior elements of the segment $[x_1, x_2]$ are simply permuted within the sum. As noted before, the origin of the different signs of $y_1$ and $y_2$ in (\ref{cm}) is illustrated in figure~\ref{fig3}: the string between $x_1$ and $x_2$ loses a mass $\mu \Delta x$ at “height” $y_2$ and gains an equal mass at “height” $y_1$.

Now, dividing (\ref{cm}) by $\Delta t = \Delta x/c$ we get the velocity $v_{CM}$ of the center of mass
\begin{equation}\label{vcm}
v_{CM}=c\frac{y_1-y_2}{x_2-x_1},
\end{equation}
from which it follows that the total transverse momentum $P$ of the string between $x_1$ and $x_2$ is given by
\begin{equation}\label{p}
P=\mu c(y_1-y_2),
\end{equation}
Incidentally, note that the total momentum depends on the difference of the displacements $y_1$ and $y_2$ at the ends. In particular, if the ends are equally displaced then the total momentum of the enclosed string would vanish! This fact alone should convince us that the transverse waves we are dealing with here are not fit to describe momentum transport along the string. For more on momentum transport see, for example,~\cite{juen}. 

To carry on with the derivation, we need to find the net external force acting on the stretch of string between $x_1$ and $x_2$. According to (\ref{force}), the transverse force acting on the left end at $x_1$ is $-Ts_1$ while that acting on the right end at $x_2$ is $Ts_2$. Here $s_1$ and $s_2$ denote the slopes of the string at $x_1$ and $x_2$ respectively. Thus, the net transverse force on the stretch of string is $T(s_2-s_1)$. Now, by Newton's second law, this must equal the time rate of change of transverse momentum $P$, which together with (\ref{p}) implies
\begin{equation}\label{n2l}
\mu c(v_1-v_2)=T(s_2-s_1),
\end{equation}
with $v_1$ and $v_2$ being the transverse velocities of the string at $x_1$ and $x_2$ respectively. Using (\ref{speedslope}) and rearranging terms in (\ref{n2l}), one obtains
\begin{equation}\label{r1}
(\mu c^2-T)(v_1-v_2)=0,
\end{equation}
Since $v_1$ and $v_2$ are totally arbitrary, we deduce that the first factor in (\ref{r1}) must vanish identically, hence (\ref{speed}).

Let's now derive (\ref{speed}) once more, this time applying the work-energy theorem to the string between $x_1$ and $x_2$ (see figure~\ref{fig3}). The kinetic energy $K$ of a small string element $\Delta x$, of mass $\mu \Delta x$, moving with velocity $v$ is given by $1/2(\mu\Delta x)v^2$. Together with (\ref{energy}) this implies that the total energy $E$ of a string element is $E = 1/2\mu v^2 \Delta x + 1/2Ts^2 \Delta x$. Referring once more to figure~\ref{fig3}), we see that the string between $x_1$ and $x_2$ loses an amount of energy equal to $E_2=1/2\mu v_2^2 \Delta x + 1/2Ts_2^2 \Delta x$ through $x_2$ and gains another amount $E_1=1/2\mu v_1^2 \Delta x + 1/2Ts_1^2 \Delta x$  through $x_1$. By the work-energy theorem, the total energy change must equal the sum of works done on the ends (we ignore internal dissipative forces). The work on the left end is given by $-F_1|\Delta y_1| =Ts_1|\Delta y_1|$ , where (\ref{force}) has been used. Similarly, the work on the right end is $-Ts_2|\Delta y_2|$ . Putting all together, we find
\begin{equation}\label{wet}
(1/2\mu v_1^2 + 1/2Ts_1^2 - 1/2\mu v_2^2 - 1/2Ts_2^2 )\Delta x=T(s_1|\Delta y_1|-s_2|\Delta y_2|).
\end{equation}
Dividing through by $\Delta x$, noting that $|\Delta y|/\Delta x = s$ and again using (\ref{speedslope}), after some rearranging of terms we arrive at
\begin{equation}\label{r2}
(\mu c^2-T)(s_1^2-s_2^2)=0,
\end{equation}
from which (\ref{speed}) follows by the arbitrariness of $s_1$ and $s_2$. It is worth noting that notwithstanding the positive slopes suggested by figure~\ref{fig3} one can easily convince oneself that (\ref{r2}) holds generically.

\section{Conclusion} 
In the paper, we present an alternative derivation of the formula (\ref{speed}) for the transverse wave speed in a string. Unlike several well-known presentations, which deal with an infinitesimal element of the string, we employ a finite element approach. It differs from other common presentations in that the considered wave pulse is of a general shape and has no inherent physical pathologies.

The pedagogical benefits of the advocated approach are twofold. First, it makes the student explicitly consider the changes in the energy and momentum distributions involved in a wave process, thus helping him/her overcome some of the misconceptions about waves commonly held by students (see e.g.~\cite{wit,cal1,cal2}). Second, it exhibits in a new light the fundamental principles of dynamics as they apply to an extended and deformable body. It is hoped that this will enrich the student’s experience and help the teacher demonstrate both the elegance and power of these principles.

Finally, a possible pedagogical caveat that arises in connection with employing the work-energy theorem is that the concept of potential energy associated with the deformation of the string is rarely ever discussed at the introductory level and the corresponding expression for the potential energy (5) is largely unfamiliar to the novice. Nonetheless, we believe that the elegance and instructiveness of the finite element approach outweigh the little price (in valuable class time) one has to pay to introduce the students to the necessary concepts.

\end{document}